\documentclass{aa}
\usepackage{graphicx}
\usepackage{txfonts}
\begin{document}

\title{Two new pulsating low-mass pre-white dwarfs or SX Phenix stars?\thanks{Visiting
  Astronomer,  Complejo  Astron\'omico   El  Leoncito  operated  under
  agreement   between   the   Consejo  Nacional   de   Investigaciones
  Cient\'{\i}ficas y  T\'ecnicas de  la Rep\'ublica Argentina  and the
  National Universities of La Plata, C\'ordoba and San Juan.}}

\author{M. A. Corti\inst{1,2}, A. Kanaan\inst{3}, A. H. C\'orsico\inst{1,4}, 
S. O. Kepler\inst{5}, L. G. Althaus\inst{1,4}, 
D. Koester\inst{6}, \and J. P. S\'anchez Arias\inst{1,4} }
\authorrunning{Corti et al.}
\titlerunning{New pulsating pre-ELM WD stars}
\institute{Grupo  de Evoluci\'on  Estelar y  Pulsaciones,  Facultad de
  Ciencias Astron\'omicas  y Geof\'{\i}sicas, Universidad  Nacional de
  La Plata, Paseo del Bosque s/n, (1900) La Plata, Argentina \and
Instituto  Argentino de  Radioastronom\'{\i}a, CCT-La  Plata, CONICET,
C.C. Nro. 5, 1984 Villa Elisa, Argentina \and
Departamento de F\'{\i}sica, Centro de Ci\^encias F\'{\i}sicas e Matem\'aticas, 
Universidade  Federal de Santa Catarina, Campus
Universit\'ario Reitor Jo\~ao David Ferreira Lima, Florian\'opolis, Brazil
\and
Instituto de Astrof\'{\i}sica La Plata, CONICET-UNLP, Paseo del Bosque
s/n, (1900) La Plata, Argentina \and
Departamento de Astronomia, Universidade Federal do Rio Grande do
Sul, Av. Bento Gon\c calves 9500, Porto Alegre 91501-970, RS, Brazil \and
Institut f\"{u}r Theoretische Physik und Astrophysik, Universit\"{a}t
Kiel, 24098, Kiel, Germany\\
\email{mariela@fcaglp.unlp.edu.ar} }

\date{Received ***; accepted ****}
\abstract{
The discovery of pulsations in low-mass stars opens an opportunity for probing their interiors and to determine their evolution, by employing the tools of asteroseismology. }
{We aim to analyze high-speed photometry of SDSSJ145847.02$+$070754.46 and SDSSJ173001.94$+$070600.25 and discover 
brightness variabilities. In order to locate these stars in the $T_{\rm eff} - \log g$ diagram we fit optical spectra (SDSS) with synthetic non-magnetic spectra derived from model atmospheres.}
{To carry out this study, we used the photometric data obtained by us for these stars with the 2.15m telescope at CASLEO, Argentina. We analyzed their light curves and we apply the Discrete Fourier Transform to determine the pulsation frequencies. Finally, we compare both stars in the $T_{\rm eff} - \log g$ diagram, with known two pre-white dwarfs, seven pulsating pre-ELM white dwarf stars, $\delta$ Scuti and SX Phe stars}
{We report the discovery of pulsations in SDSSJ145847.02$+$070754.46 and SDSSJ173001.94$+$070600.25. We determine their effective temperature and surface gravity to be $T_{\rm  eff}$ = 7\,972 $\pm$ 200 K, $\log g$ = 4.25 $\pm$ 0.5 and $T_{\rm  eff}$ = 7\,925 $\pm$ 200 K, $\log g$ =  4.25 $\pm$ 0.5, respectively. With these parameters these new pulsating low-mass stars can be identified with either ELM white dwarfs (with $\sim 0.17\,M_{\sun}$) or more massive SX Phe stars. We identified pulsation periods of 3\,278.7 and 1\,633.9 s for SDSSJ145847.02$+$070754.46 and a pulsation period of 3\,367.1 s for SDSSJ173001.94$+$070600.25. These two new objects together with those of \citet{max13, max14} indicate the possible  existence of a new instability domain towards the late stages of evolution of low-mass white dwarf stars, although their identification with SX Phe stars cannot be discarded.}
{} 
\maketitle
%
\section{Introduction} 
\label{sec:intro}

Low-mass ($M_{\star}/M_{\sun} \lesssim 0.45$) white dwarfs (WDs) are
thought to be the outcome of strong mass-loss events at the red giant
branch stage of low-mass stars in binary systems before the He flash
onset \citep{alt10}. Since the He flash does not occur, their cores
must be made of He. This is in contrast to the case of average-mass
($M_{\star} \sim 0.6\,M_{\sun}$) WDs which are thought to have C/O
cores. Binary evolution is the most likely origin for the so-called
extremely low-mass (ELM) WDs, which have masses below $\sim 0.18-0.20\,M_{\sun}$. 
It is well known \citep{drie98} that ELM WDs are expected to harbor very thick H envelopes able to
sustain residual H nuclear burning via $pp-$chain, leading to markedly
long evolutionary timescales. Numerous low-mass WDs, including ELM
WDs, have been detected through the ELM survey and the SPY and WASP
surveys \citep{koes09, brow10, brow12, brow13, max11, kil11, kil12,
  gia15}. The interest in low-mass WDs has been greatly boosted by the
discovery that some of them  pulsate \citep[ELMVs,][]{her12, her13a,
  her13b, kil15, bel15a}, which constitutes an unprecedented
opportunity for probing their interiors and eventually to test their
formation channels by employing the tools of
asteroseismology. 

High-frequency pulsations in a precursor of a low-mass WD star
component of an eclipsing binary system were reported by
\citet{max13}. The object analyzed by those authors, 1SWASP
J024743.37$-$251549.2B (hereinafter WASPJ0247$-$25B, $\log  g= 4.58$,
$T_{\rm  eff}= 11\,380$ K, $M_{\star}= 0.186\,M_{\sun}$) showed
variability with periods in the range 380$-$420 s, likely due to
a mixture of radial ($\ell= 0$) and nonradial ($\ell \geq 1$) $p$
modes.  Theoretical models predict that this star is evolving to
higher effective temperatures at nearly constant luminosity prior to
becoming a low-mass WD. This could be the first member of a new class
of pulsating stars that are the precursors of low-mass WDs. A second
object of this type, 1SWASP J162842.31$+$101416.7B (hereinafter
WASPJ1628$+$10B, $\log g= 4.49$, $T_{\rm eff}= 9\,200$ K, $M_{\star}=
0.135\,M_{\sun}$) was discovered by \cite{max14} in other eclipsing
binary system, showing high-frequency signals likely to be due to
pulsations similar to those seen in WASPJ0247$-$25B.
However, additional photometry is required to confirm that the
high-frequency variations are due to pulsations in the pre-WD star,
and not produced by the companion star, a normal A2V star showing
$\delta$ Scuti-type pulsations. Shortly after the discovery of the
short-period oscillations in WASPJ0247$-$25B,
\citet{jef13} explored the pulsation stability of radial modes of
low-mass WD models considering a range of envelope chemical
composition, effective temperature and luminosity.  \citet{jef13}
identified the instability boundaries associated with radial modes
characterized by low-to-high radial orders, and showed that they are
very sensitive to composition. These authors found that the excitation
of modes is by the $\kappa-$mechanism operating mainly in the second
He-ionization zone, provided that the driving region is depleted in H
($0.2 \lesssim X_{\rm H} \lesssim 0.3$).

It is worth noting that the domain of the precursors of ELM WDs in the $T_{\rm  eff}-\log g$ 
plane could be coincident with that corresponding to the pulsating $\delta$ Scuti or SX Phe stars.
It is not possible to distinguish between both classes of pulsating stars by amplitude and number of frequencies. 
There are several SX Phe stars with characteristics
similar to the high-amplitude $\delta$ Scuti stars except that field SX Phe have
high proper motions and low metallicity \citep{fau10}.
SX Phe stars probably belong to the thick disc or
halo and may therefore be considered to be in an advanced stage
of evolution, while the $\delta$ Scuti are main sequence stars. 
The SX Phe stars that are not field stars are to be found in globular clusters \citep{bal12}.

The members of both SX Phe and $\delta$ Scuti classes of variable stars have $1.5-2.3\,M_{\sun}$ and 
exhibit pulsations due to low-order radial  and nonradial
$p$ modes with periods between 12 min to 10 hours \citep{cat15}. Most of SX Phe and $\delta$ Scuti stars have 6\,000 $< T_{\rm  eff} <$ 9\,000 K and the logarithm of gravities around 4.0. However, some of these stars are hotter than the average, thus populating the same region of the $T_{\rm  eff}-\log g$ diagram where precursors of ELM WDs could be located.

In this paper, we report the discovery of two new pulsating SDSSJ173001.94$+$070600.25 ($T_{\rm  eff}$ = 7\,972 $\pm$ 200 K, $\log g$ = 4.25 $\pm$ 0.5) and 
SDSSJ145847.02$+$070754.46 ($T_{\rm eff}$ = 7\,925 $\pm$ 200 K, $\log  g$ = 4.25 $\pm$ 0.5), that could be precursors of ELM WDs or instead, SX Phe stars.
In Sections \ref{sec:data} and \ref{sec:analysis}, we present the observational material and the analysis of the data, respectively. In Section \ref{sec:discussion} 
we discuss the evolutionary status of these new pulsating stars and give a brief conclusion.

\section{Data}
\label{sec:data}

The observational material consists of high-speed photometry of the
J1458$+$0707 and J1730$+$0706 stars obtained in 2015 during the
  nights of April 24$^{th}$ to 26$^{th}$. We used a TEK
1024$\times$1024 CCD with pixel size of 24 $\mu$m, binned by a
factor 2, attached to the 2.15$-$m telescope at Complejo Astron\'omico
El Leoncito (CASLEO)\footnote{Operated under agreement between
  CONICET, SeCyT, and the Universities of La Plata, C\'ordoba and San
  Juan, Argentina.} in San Juan, Argentina. Observations were obtained
through a red cutoff 3 mm BG40 filter to reduce sky background.

We obtained a total of 156 frames of J1458$+$0707
during the two first nights (119 the first and 37 the second night)
and 89 frames of J1730$+$0706 during the last night. The exposure
time for each image was 120 s, the seeing averaged 2 seconds of arc and the
second night was cut short by clouds. Bias, dark and flat-field frames were
obtained every night. All images were obtained by employing the data
  acquisition program ROPER and these were analyzed applying aperture
photometry with the external IRAF package $ccd_{hsp}$ \citep{kana02}).

\section{Analysis}
\label{sec:analysis}

SDSSJ1458$+$0707 and SDSSJ1730$+$0706 were classified as A stars by the {\small ELODIE}
pipeline of DR12 of SDSS. As part of our pulsating ELM candidate selection program we fitted the optical spectra to pure
hydrogen local thermodynamic equilibrium (LTE) grids of synthetic
non-magnetic spectra derived from model atmospheres \citep{koes10}. The model grid uses the ML2$/\alpha= 0.8$ approximation.
Our grid covers effective temperature 
and gravity values in the ranges $6\,000~K \leq T_\mathrm{eff} \leq 100\,000$~K, and $3.75 \leq \log g \leq 10.0$, respectively, covering 
main sequence, sudbwarfs and white dwarf spectra. 

We fitted the SDSS spectral lines with models that work well in the white dwarf regime above $\log g \sim$ 5.50.  
To adjust the spectral lines with lower gravity models it becomes increasingly difficult due to the degeneracy of 
the parameters becomes very significant. 
A change in surface gravity can almost completely be compensated by a change in temperature. The moderate resolution of 
the SDSS spectra is another problem. The higher Balmer lines become narrower at low surface gravity and are thus strongly
affected by the instrumental profile. On the other hand, the SDSS photometry is quite useful to lift this degeneracy, mainly
because of the highly gravity sensitive Balmer jump measured by the u-g color. We therefore started the analysis with a fit of 
the photometry with theoretical values. The possible reddening is not a severe problem, the objects are very far outside the galactic
plane. Assuming a fixed $\log g$ between 3.75 and 4.75 the best fitting temperature is almost constant. As a rather conservative 
estimate we use a result of the photometry fitting the values in Table \ref{tab:periods}. 
The calcium abundance was determined from a comparison of theoretical line strengths with the observed spectra. 
The errors include also the complete range of errors of the parameters.

\subsection{SDSSJ1458$+$0707}

\begin{figure*}[!ht]
\centering
\includegraphics[width=0.8\textwidth,]{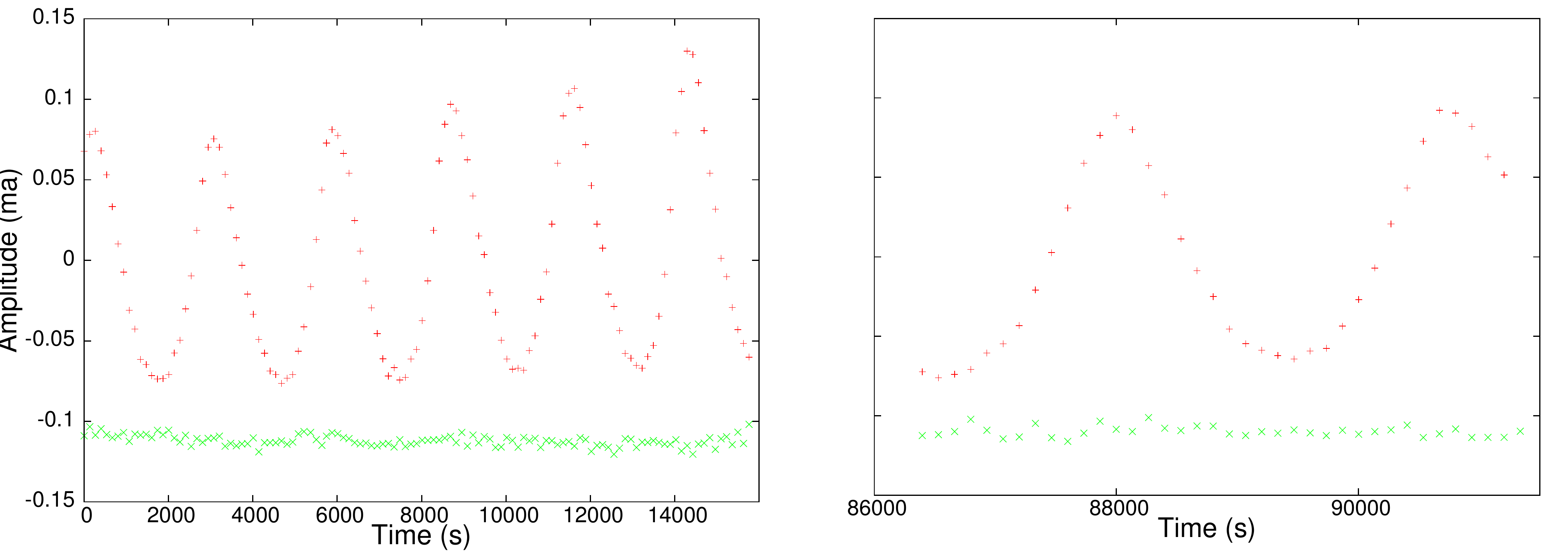}
\caption{Top curve: high-speed photometry of J1458$+$0707 star over data points obtained during two consecutive nights. Bottom curve: time variations of the brightest comparison star in the field over the same period (see text and Table \ref{tab:periods}).} 

\label{fig:cl1}
\end{figure*}

To build the light curve shown in Figure \ref{fig:cl1} we determined the aperture which provides the light curves with minimum scatter. The optimum aperture radius was 12 pixels. The sky annulus used for sky subtraction had an inner radius of 20 pixels and 6 pixels width.  
The light curve for the pulsating star and a comparison star of similar intensity are shown in Figure \ref{fig:cl1}. Both stars were divided by the same comparison star, 0.4 magnitudes brighter, to correct for atmospheric transparency fluctuations. Each point in the light curves was normalized by the average of all light curve points. The variations visible in the comparison star light curve are attributed to noise.

\begin{figure}[!ht]
\centering
\includegraphics[width=0.4\textwidth, angle=0]{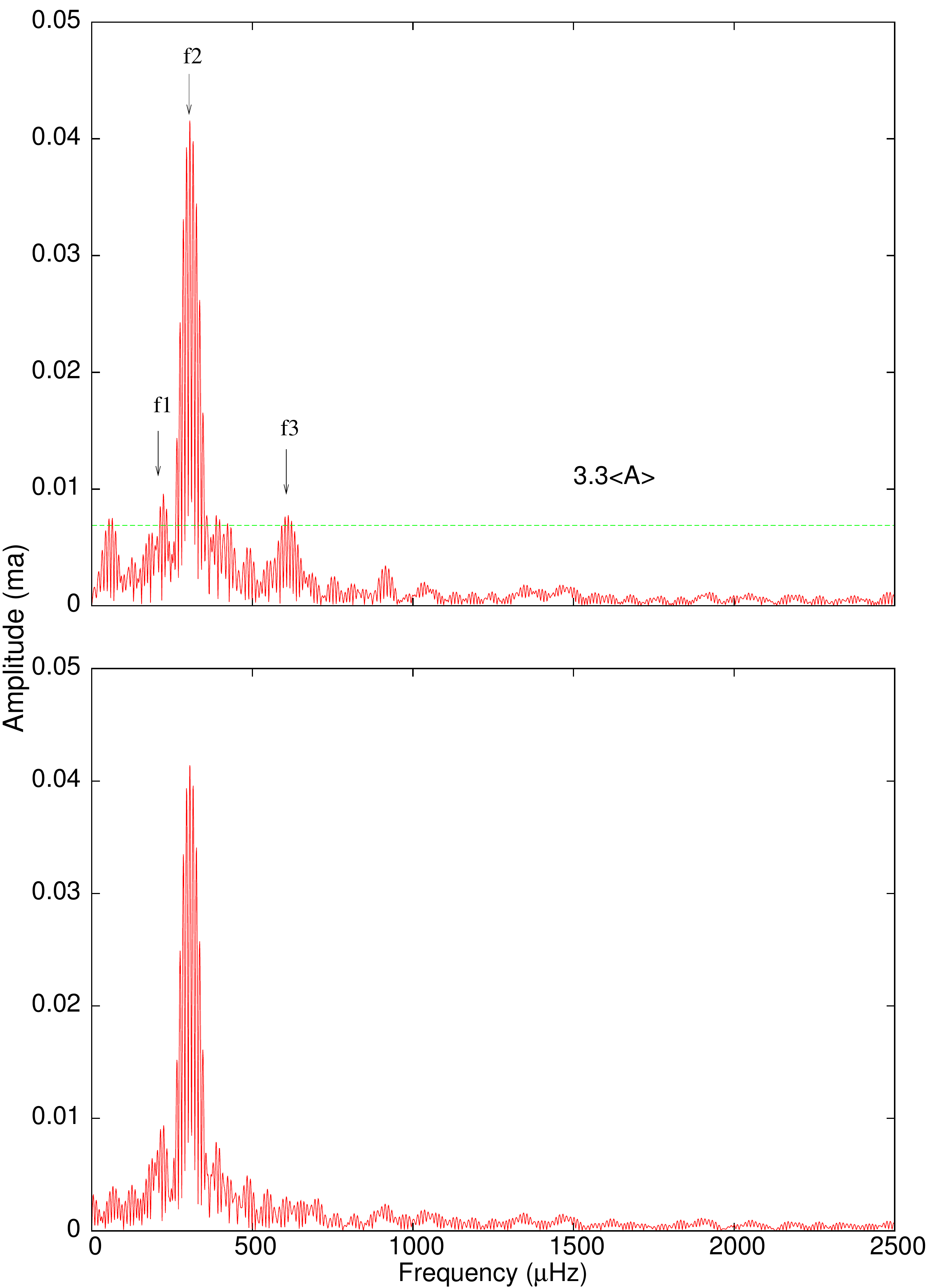}
\caption{Top panel shows the discrete Fourier Transform of the light curve of J1458$+$0707. The horizontal line shows 3.3 times the noise level. There are three pulsation frequencies (see text and Table \ref{tab:periods}). The bottom panel shows the spectral window with more intense emission peak at 305 $\mu$Hz.}
\label{fig:cl1TF}
\end{figure}

We computed the Fourier Transform of the two nights' data joining the two light curves after correcting the timings to Barycentric Julian Ephemeris Date (IAU SOFA).  To check for sampling artifacts we computed a spectral window as well.  Figure \ref{fig:cl1TF}  shows the Fourier Transform and the spectral window.  The main peak in the FT is labelled $f_2$ and corresponds to the variations clearly seen in the light curve.  The small peak $f_1$ has an analog in the spectral window at the same frequency and same intensity relative to the $f_2$, we dismiss $f_1$ as an alias peak. Another frequency with intensity above $ 3.3 \langle {\rm A} \rangle$ noise limit \citep{kep93} is $f_3$. It is probably an $f_2$ harmonic because it is almost twice $f_2$.

\subsection{SDSSJ1730$+$0706}
\begin{figure}[!ht]
\centering
\includegraphics[width=0.3\textwidth,angle=-90]{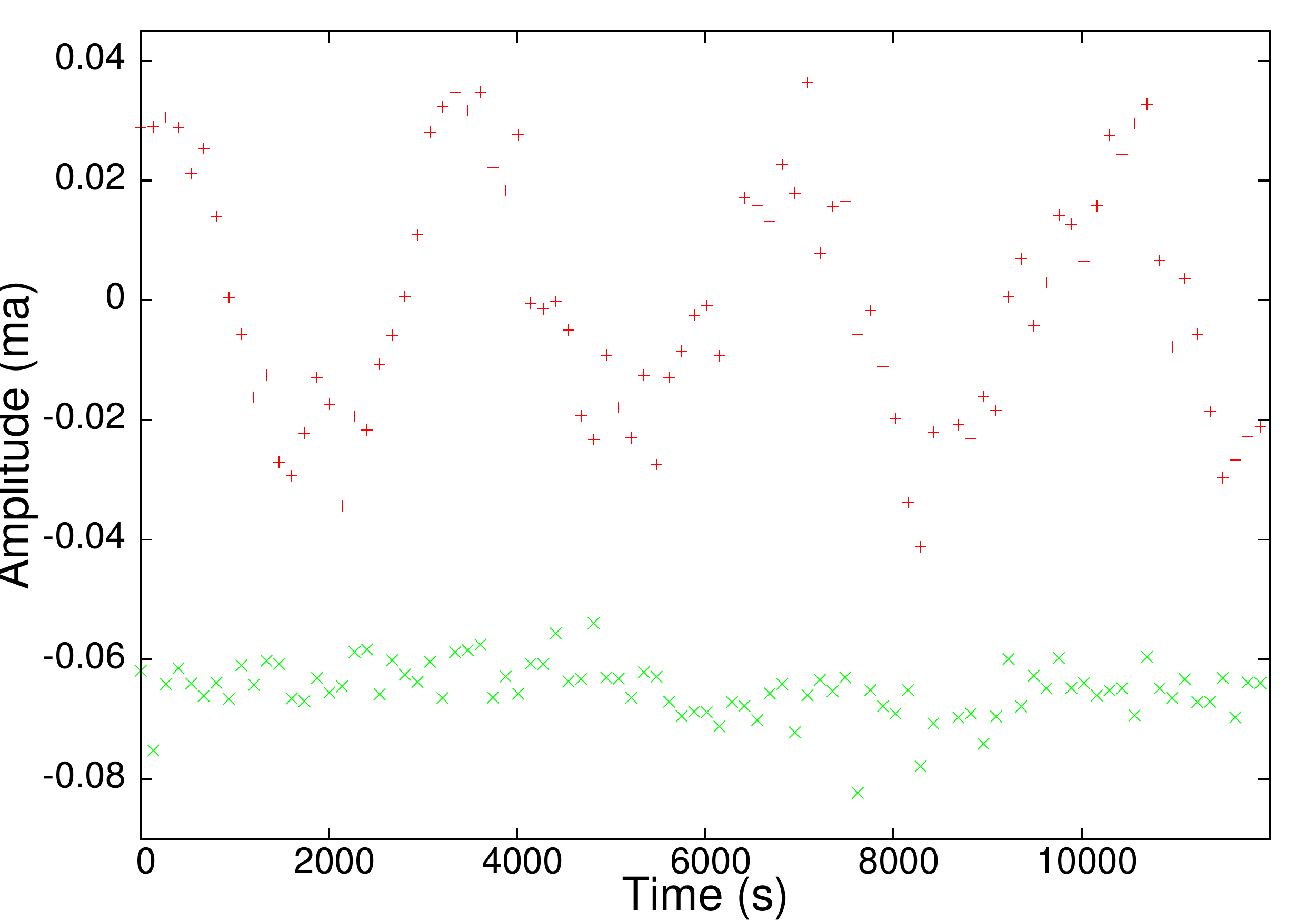}
\caption{Top curve: high-speed photometry of J1730$+$0706 star over data points obtained during one night. Bottom curve: 
time variations of the brightest comparison star in the field over the same period (see text and Table \ref{tab:periods}).} 

\label{fig:cl2}
\end{figure}

This star and its comparison stars were analyzed in the same manner as J1458$+$0707. The top light curve shown in Figure \ref{fig:cl2} corresponds to J1730$+$0706 divided by a comparison star 1 magnitude brighter.  
The bottom light curve was obtained for the brightest comparison stars (1.5 magnitude brighter than the target star) divided by the same comparison star used for the target object. The peak to peak variations in this last light curve reach 0.01 ma, totally attributable to noise.  

\begin{figure}[!ht]
\centering
\includegraphics[width=0.45\textwidth,angle=0]{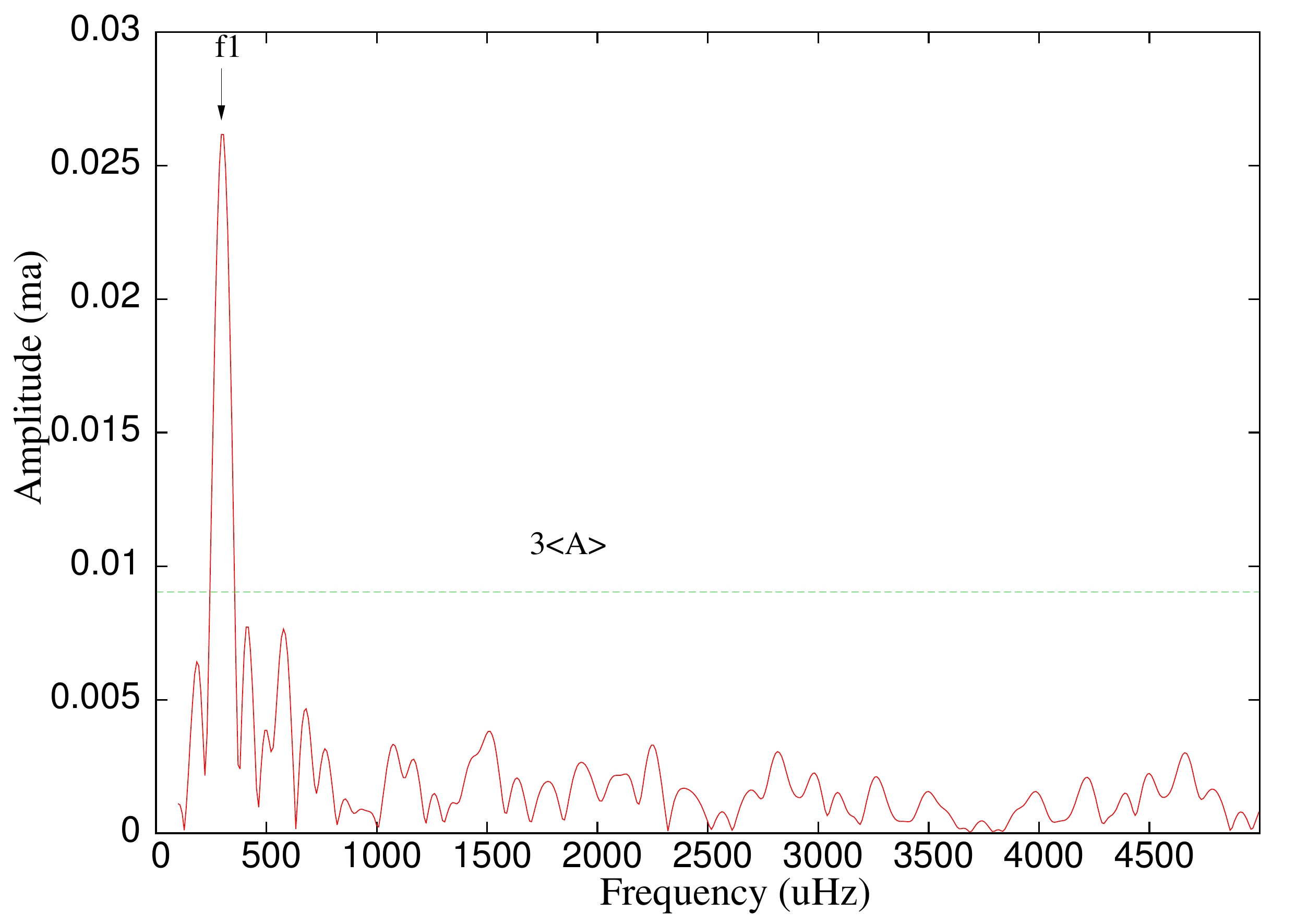}
\caption{Discrete Fourier Transform of the J1730$+$0706 light curve. The horizontal line shows three times the noise level. There is one pulsation frequency (see text and Table \ref{tab:periods}).}
\label{fig:cl2TF}
\end{figure}

We employed the same Fourier analysis technique as for J1458$+$0707, the Fourier transform for this object is shown in Figure \ref{fig:cl2TF}. with an average modulation amplitude $<A> = 3.0$ \, mma \citep{kep93}. The parameters for the only significant peak are shown in Table \ref{tab:periods}.

\begin{table*}
\fontsize{10} {14pt}\selectfont
\caption{Data of pulsating stars}
\centering
\begin{tabular}{cccccccccc}
\hline\hline
ID & Period & Frequency & Amplitude & $T_{\rm  eff}$ & $\log g$ &  [Ca/H] & Z & $\mu^{\triangle}$ \\
SDSS     & (s)       &  ($\mu$Hz) & (ma)    &  (k)  &     &  & (pc) & ($mas\, yr^{-1}$)  \\ 
\hline
 J$145847.02+070754.46$  & 3278.7 &  f2=305 &  0.042 & 7\,925 $\pm$ 200 & 4.25 $\pm$  0.5 &  -7.00 $\pm$ 0.25 & 2559 & 55 $\pm$ 4 \\
                            &  1633.9 &  f3=612 &  0.008  & & & & & \\
 J$173001.94+070600.25$  & 3367.1 &  f1=297 & 0.026 &  7\,972 $\pm$ 200 &  4.25 $\pm$  0.5 & -6.55 $\pm$ 0.25 & 1745 &  12 $\pm$ 5 \\
 
 \hline
 \label{tab:periods}
\end{tabular}
\tablefoot{ 
\tablefoottext{$^{\triangle}$}{\citet{ucac4}}}
\vspace{-0.05cm}
\end{table*}

\section{Discussion and Conclusions}
\label{sec:discussion}

\begin{figure}[!ht]
\centering
\includegraphics[width=0.45\textwidth, angle=0]{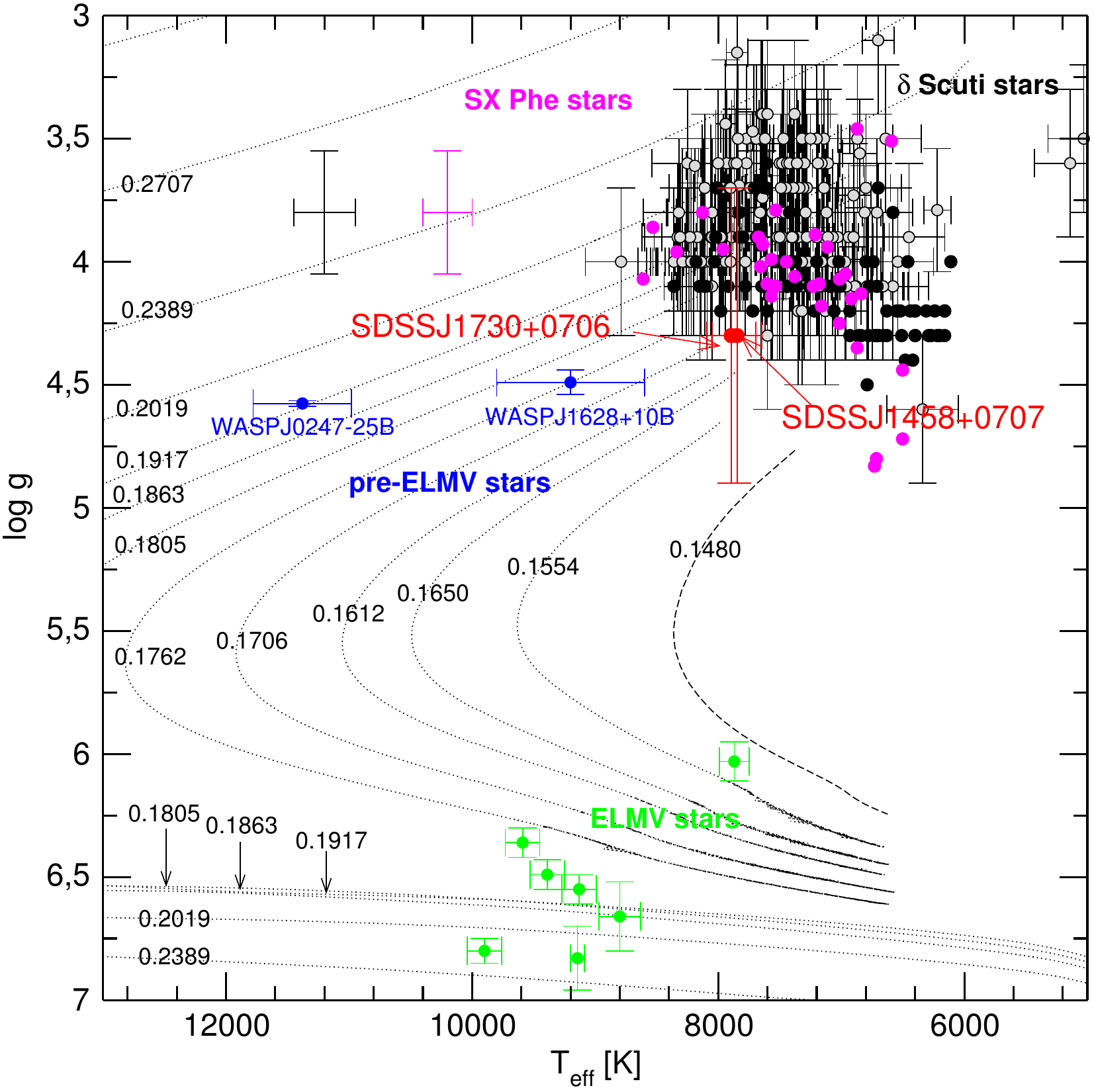}

\caption{Location of the stars SDSSJ1730$+$0706 and SDSSJ1458$+$0707 in the $T_{\rm eff}-\log g$ plane, with samples of other classes of variable stars (see the text). The crosses on the left side indicate the error of the black and pink dots. The extremely short$-$lived stages during the CNO flashes episodes of $M_{\star} > 0.1762\,M_{\sun}$ are not shown in the figure.}
\label{fig:HR}
\end{figure}

The  seven pulsating  ELM  WD  stars known  \citep{her12, her13a, her13b, 
kil15, bel15a} have effective temperatures between $\sim 10\,000$ K and $\sim 7500$ K, and
$6 < \log g < 7$.  The
location  of  these stars  in  the  $T_{\rm  eff}-\log g$  diagram  is
depicted  in Fig. \ref{fig:HR} as light green dots. Also shown are the
evolutionary   tracks  of   low-mass  He-core   WDs  as   computed  by
\citet{alt13}   (dotted black  lines)   in   the   range  of   masses
$0.1554-0.2707\,M_{\sun}$. For the sake of completeness, we include also a $0.148\,M_{\sun}$ evolutionary track computed by 
\citet{sere01}  (dashed  black   line). In addition, we show with grey and black dots the location of a sample of $\delta$ Scuti stars extracted from \citet{uyt11} and \citet{brad15}, respectively, and the location of a sample of SX Phe stars with pink dots, extracted from the \citet{bal12}.
Finally,  the locations of SDSSJ1730$+$0706 and SDSSJ1458$+$0707 are marked with red
dots, along  with   the   already  known   pulsating  pre-WD   stars
WASPJ0247$-$25B and WASPJ1628$+$10B  \citep{max13, max14} which are marked with blue dots.
Clearly, given  the very low surface gravities
characterizing SDSSJ1730$+$0706  and SDSSJ1458$+$0707 ($\log  g \sim 4.3$),
these two stars cannot be identified with low-mass WDs.
Instead, they could be associated to either low-mass pre-WD stars with stellar masses between $\sim 0.17$ and $\sim  0.20\,M_{\sun}$ or to the hottest known $\delta$ Scuti or SX Phe stars. 
 
If the stars presented in this work are indeed precursors of low-mass WDs, we can trace their following evolution. 
Both stars would be evolving to hotter effective temperatures directly to its terminal cooling track. We would conclude that these
new pulsating stars are of  the same type as the  pulsating pre-WDs WASPJ0247$-$25B  and WASPJ1628$+$10B. 

We note, however,  that there  exists some  differences concerning  the
observed pulsation  properties.  Indeed, SDSSJ1730$+$0706  and
SDSSJ1458$+$0707 pulsate  with periods  in the  range 1634-3367 s,
longer than those exhibited  by  WASPJ0247$-$25B  and WASPJ1628$+$10B,
which show periods in the range $400-800$ s  \citep{max13, max14}.
 Also, it is apparent that  the  amplitudes observed  in
  SDSSJ1730$+$0706 and  SDSSJ1458$+$0707 (up to $\sim 40$ mmag)  
  are larger than those detected in WASPJ0247$-$25B  and WASPJ1628$+$10B  
  (of the  order of a   few mmag). According  to the
theoretical analysis  of \citet{jef13}, at  least the   pulsations of
WASPJ0247$-$25B  should  be  due   to  radial  modes  of  high  radial
order. Instead, the long periods observed in SDSSJ1730$+$0706 and
SDSSJ1458$+$0707 are likely due to high-order nonradial $g$  modes.
The question of why short periods typical of $p$ modes (or
radial modes) are not detected in SDSSJ1730$+$0706 and
SDSSJ1458$+$0707, and vice versa, why no long periods characteristic
of $g$ modes are observed in WASPJ0247$-$25B  and WASPJ1628$+$10B,
could be answered through a full  (radial and  nonradial) linear
pulsation stability  analysis. It  will  be  addressed in  a  future
investigation. Concerning the contrast in the amplitudes of $g$
modes of  SDSSJ1730$+$0706 and SDSSJ1458$+$0707 as compared with
those of the $p$ modes exhibited  by  WASPJ0247$-$25B  and
WASPJ1628$+$10B, a definitive answer could be given by non-linear
pulsation calculations, which are out the scope of this paper.

If instead, these stars with metallicity lower than solar metallicity and with proper motion of the stars in the galaxy halo 
were SX Phe stars, they would populate the hottest portion of the SX Phe instability strip. Reversing the argument,
it cannot be discarded, by inspecting Fig \ref{fig:HR}, the possibility that some of the hottest known SX Phe stars would be otherwise low-mass pre-WD stars.

\begin{acknowledgements}
M.A.C acknowledge support from CONICET (PIP 112-201201-00226). A.K. and S.O.K. acknowledge CNPq support. D.K. received support from programme Science without Borders, MCIT/MEC-Brazil. We thank to Dr. Costa Jos\'e Eduardo to allow us to use his DFT program. 
We wish to thank the anonymous referee for the suggestions and comments that improved the original version of this work.
\end{acknowledgements}

\bibliographystyle{aa}  
\bibliography{bibliojulio2015}
\end{document}